\documentclass[twocolumn,prl, showpacs]{revtex4}
\usepackage{amssymb}
\usepackage{amsmath}
\usepackage{bbm}
\usepackage{graphicx}

\begin{document}

\author{Philipp Werner}
\author{Andrew J. Millis}
\affiliation{Columbia University, 538 West, 120th Street, New York, NY 10027, USA}
\title{Efficient DMFT-simulation of the Holstein-Hubbard Model}

\date{August 16, 2007}

\hyphenation{}

\begin{abstract}
We present a method for solving impurity models with electron-phonon coupling, 
which treats the phonons efficiently and without approximations. The algorithm is 
applied to the Holstein-Hubbard model in the dynamical mean field approximation, 
where it allows access to strong interactions, very low temperatures and arbitrary fillings. 
We show that a renormalized Migdal-Eliashberg theory provides a reasonlable description of the phonon contribution to the electronic self energy in strongly doped systems, but fails if the quasiparticle energy becomes of order of the phonon frequency.
\end{abstract}


\maketitle
 
Electron-lattice interactions play a fundamental role in the physics of metals. In conventional 
(``weakly correlated") materials such as Al or MgB$_2$, phonons are the dominant source of 
inelastic scattering and provide the pair binding which leads to superconductivity. In these 
materials, physical understanding is greatly aided by the Migdal-Eliashberg theory, which 
justifies the use of density functional band theory methods to estimate the electron-phonon
coupling constants and perturbative methods to estimate electron mass renormalizations, 
lifetimes and pairing strengths.  A key result of the Migdal-Eliashberg theory is that 
the effect of the electrons on the phonons is a renormalization of the phonon 
oscillator frequency  $\omega_0$,  while the effect of the phonons on the electrons
is an increase in the electron effective mass. The mass  increase ``turns on"
at frequencies below the renormalized phonon frequency. 

In unconventional (``strongly correlated") materials such as the high temperature 
superconductors \cite{Lanzara01}, the fullerenes \cite{Gunnarsson97, Capone02}, or the 
colossal magnetoresistance rare-earth manganites \cite{Millis95, Yamasaki06} the situation is 
less clear. Many experiments suggest that electron-phonon effects are important. For example,
in high $T_c$ superconductors changes attributed to the electron-phonon coupling are observed in
the electronic dispersion \cite{Lanzara01} at energies
of the order of typical optical phonon frequencies.  However, there is as yet no
clear theoretical basis for interpreting these or related data in strongly correlated
materials. While there has been extensive and important work on static properties 
of models involving electron-electron and electron-phonon coupling 
\cite{Freericks,Tezuka05}, on Fermi-liquid and effective field theory based approaches \cite{Levin,Grilli06}, on the Holstein model \cite{Benedetti98}
and on the (bi-)polaron problem (one or two interacting electrons)
\cite{Bonca00,MischenkoNagaosa}
less is known about the dynamical consequences of the electron-phonon 
interaction in strongly correlated materials. 

A  way forward is provided by dynamical mean field theory (DMFT), 
a powerful, non-perturbative tool to study the properties of 
strongly correlated systems, which has 
provided considerable insight into the correlation induced 
metal-insulator (Mott) transition \cite{Georges96}. Progress in using  
dynamical mean field methods to study phonons coupled
to strongly correlated electrons 
\cite{Deppeler02,Koller04, Capone04}
has been hampered
by the mismatch in energy scales between the electronic and lattice parts of the problem
and by the difficulty of providing a quantitative treatment of the low temperature
properties of strongly correlated materials even in the absence of electron-phonon coupling.
In this paper we introduce a new
method which resolves these problems and allows the DMFT simulation of 
important classes of models at essentially the same 
computational expense as the corresponding models without coupling to phonons. 

We consider the Holstein-Hubbard Hamiltonian 
\begin{eqnarray}
H&=&-\sum_{i,\delta,\sigma}t(\delta)c^\dagger_{i+\delta,\sigma}c_{i,\sigma}+\sum_i [Un_{i,\uparrow}n_{i,\downarrow}-\mu(n_{i,\uparrow}+n_{i,\downarrow})]\nonumber\\
&&+\lambda\sum_i (b^\dagger_i+b_i)(n_{i,\uparrow}+n_{i,\downarrow}-1)+\omega_0\sum_i b^\dagger_i b_i,
\label{H}
\end{eqnarray}
where $U$ denotes the on-site repulsion, $\mu$ the chemical potential of the electrons with creation operators $c^\dagger_\sigma$ and density operators $n_\sigma$, $b^\dagger$ the creation operator for Einstein phonons of frequency $\omega_0$, and the electron-phonon coupling is $\lambda$.  

The single-site DMFT approximation \cite{Georges96} reduces
the problem to the self-consistent solution of a quantum impurity model specified by the Hamiltonian
$H_\text{QI} = H_\text{loc}+H_\text{hyb}+H_\text{bath}$. Here, 
the local term is 
\begin{eqnarray}
H_\text{loc} &=& -\mu(n_\uparrow+n_\downarrow)+U n_\uparrow n_\downarrow\nonumber\\
&& +\lambda (n_\uparrow+n_\downarrow-1) (b^\dagger+b)+\omega_0b^\dagger b, 
\label{H_loc}
\end{eqnarray}
and the impurity-bath mixing and bath Hamiltonians are
$H_\text{hyb} =\sum_{p, \sigma} (V_{p, \sigma} c^\dagger_\sigma a_{p, \sigma} + V^*_{p, \sigma} c_\sigma a^\dagger_{p, \sigma})$ and 
$H_\text{bath} =\sum_{p, \sigma} \epsilon_p a^\dagger_{p, \sigma} a_{p, \sigma}$. 
The parameters $V_{p, \sigma}$ and $\epsilon_p$ are determined by a self-consistency equation.

In the absence of electron-phonon coupling, $H_\text{loc}$ has a small Hilbert space
and one energy scale  ($U$). 
Adding an electron-phonon coupling introduces a new energy scale, the phonon frequency $\omega_0$,
and requires keeping track of the bosonic sector of the Hilbert space (with an infinite number of states). Previous approaches to the 
problem have involved either treating the bosons semiclassically 
\cite{Deppeler02,Millis96a}
or truncating the boson Hilbert space, 
retaining only a  finite number of boson states 
\cite{Koller04,Capone04}. 
The semiclassical approach 
cannot account for quantal phonon effects such as electronic
mass renormalization or superconductivity, while treating even a truncated boson Hilbert space directly 
is computationally expensive. 

In Refs.~\cite{Werner05, Werner06} we have shown  that the stochastic sampling of
a diagrammatic expansion of the partition function in powers of the 
impurity-bath hybridization term $H_\text{hyb}$ 
leads to a highly efficient \cite{Gull06}  impurity solver 
for purely electronic models.
After tracing out the bath states $a_{p, \sigma}$, the weight of a Monte 
Carlo configuration corresponding to a perturbation order $n$ ($n$ creation 
operators $c^\dagger_\sigma(\tau_\sigma)$ and $n$ annihilation operators 
$c_\sigma(\tau'_\sigma)$) can be expressed as
\begin{eqnarray}
&&w(\{O_i(\tau_i)\}) = Tr_c \Big\langle  T_\tau e^{-\int_0^\beta d\tau H_\text{loc}(\tau)} O_{2n}(\tau_{2n})\ldots\nonumber\\
&&\hspace{5mm}\ldots O_2(\tau_2)O_{1} (\tau_{1}) \Big\rangle_bd\tau_1\ldots d\tau_{2n}\prod_\sigma (\det M_\sigma^{-1})s_\sigma, \hspace{5mm}
\label{weight}
\end{eqnarray}
with $O_i(\tau_i)$ the (time ordered) creation and annihilation operators for spin up or down
electrons on the impurity site. The matrix elements $(M_\sigma^{-1})_{i,j}=F_\sigma(\tau'_{\sigma,i}-\tau_{\sigma,j})$ are determined by the $V_{p,\sigma}$ and $\epsilon_p$ through the hybridization functions 
$F_\sigma(-i\omega_n)=\sum_p \frac{|V_{p,\sigma}|^2}{i\omega_n-\epsilon_p}$ \cite{Werner06}.
The sign $s_\sigma$ is $1$ if the $\sigma$-spin operator with the lowest time argument is a creation operator and $-1$ otherwise.  

To evaluate $\left<\cdots\right>_b$ we use a Lang-Firsov \cite{Lang62} transformation.
Defining operators $X=(b^\dagger +b)/\sqrt{2}$ and $P=(b^\dagger -b)/i\sqrt{2}$,
the unitary transformation specified by $e^{iPX_0}$ shifts $X$ by $X_0=(\sqrt{2}\lambda/\omega_0)(n_\uparrow+n_\downarrow-1)$ so that  
\begin{eqnarray}
\tilde H_\text{loc} &=& e^{iPX_0}H_\text{loc} e^{-iPX_0}\nonumber\\
&=& -\tilde \mu(\tilde n_\uparrow+\tilde n_\downarrow)+\tilde U \tilde n_\uparrow \tilde n_\downarrow+\frac{\omega_0}{2}(X^2+P^2) 
\label{H_loc_transf}
\end{eqnarray}
has no explicit electron-phonon coupling. ${\tilde H}_\text{loc}$ is of the Hubbard form but
with modified chemical potential and interaction strength:
$\tilde \mu=\mu-\lambda^2/\omega_0,$  $\tilde U=U-2\lambda^2/\omega_0.$
Also, the  electron creation and annihilation operators are transformed according to
$\tilde c^\dagger_\sigma = e^{\frac{\lambda}{\omega_0}(b^\dagger-b)}c^\dagger_\sigma$ and 
$\tilde c_\sigma = e^{-\frac{\lambda}{\omega_0}(b^\dagger-b)}c_\sigma$.
The phonon contribution $w_b$ to the weight~(\ref{weight}) is
$w_b(\{O_i(\tau_i)\})=\left<e^{s_{2n}A(\tau_{2n})}\cdots e^{s_1A(\tau_1)}\right>_b$
with $0\le \tau_1<\ldots <\tau_{2n}<\beta$, $s_i=1$ $(-1)$ if the $n^\text{th}$ operator is a creation (annihilation) operator and  $A(\tau)=\frac{\lambda}{\omega_0}(e^{\omega_0\tau}b^\dagger-e^{-\omega_0\tau}b)$. This expectation value, taken in the thermal state of free bosons, evaluates to 
\begin{eqnarray}
&&w_b(\{O_i(\tau_i)\})=\exp\bigg[-\frac{\lambda^2/\omega_0^2}{e^{\beta \omega_0}-1}\Big(n \big(e^{\beta \omega_0}+1\big)\nonumber\\&&\hspace{5mm}+\sum_{2n\geq i>j\geq 1} s_is_j\big\{e^{\omega_0(\beta-(\tau_i-\tau_j))}+e^{\omega_0(\tau_i-\tau_j)}\big\}\Big)\bigg],\hspace{5mm}
\label{wb}
\end{eqnarray}
and the weight (\ref{weight}) becomes the product
\begin{equation}
w(\{O_i(\tau_i)\})=w_b(\{O_i(\tau_i)\})\tilde w_\text{Hubbard}(\{O_i(\tau_i)\}),
\end{equation}
where $\tilde w_\text{Hubbard}$ is the weight of a corresponding configuration in the Hubbard impurity model (without phonons, but modified parameters $\tilde U$ and $\tilde \mu$).
The Holstein-Hubbard model can thus be simulated without truncation at 
an expense comparable to the Hubbard model.  Superconducting phases can be simulated in
the same way, using the formalism outlined in Ref.~\cite{Georges96}; the only difference
is that the determinant in Eq.~(\ref{weight}) can no longer be factorized into spin components. 


We have applied our method to the Holstein-Hubbard model with a
semi-circular density of states of bandwidth $4t$ and $\omega_0/t=0.2$. 
For reasons of space, we restrict our attention to non-superconducting phases. 
%
%
\begin{figure}[t]
\begin{center}
\includegraphics[angle=-90, width=0.9\columnwidth]{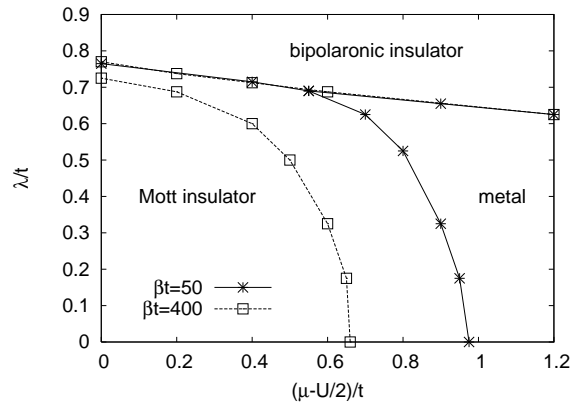}
\caption{Phase diagram for $\omega_0=0.2t$,  $\beta t=50,400$ and $U/t=6$ in the plane of chemical potential $\mu$ and phonon coupling strength $\lambda$. Half-filling corresponds to $\mu=U/2$. In the metallic phase, the filling increases with increasing $\mu$, $\lambda$. 
}
\label{lambda_mu}
\end{center}
\end{figure}
Figure~\ref{lambda_mu} shows 
the phase diagram at $U/t=6$ in the space of chemical potential and phonon coupling. Our results at half-filling are in good agreement with those computed  by NRG \cite{Koller04}.
If the chemical potential is increased at fixed electron-phonon coupling, a transition to a metal occurs. 
The shift in critical $\mu$ with temperature arises from the entropic stabilization of the insulating phase and
is roughly $\lambda$-independent. The transition is first order at $T>0$
(with a jump in density), but apparently becomes continuous as $T\rightarrow 0$. 
The critical chemical potential decreases as the electron-phonon coupling
is increased. This behavior is physically expected: increasing the phonon coupling reduces the
effective interaction $\tilde U$ and so the magnitude of the insulating
gap. We have confirmed this by analysis of the Green function.
Our finding appears to contradict Ref.~\cite{Capone04} which stated that the 
electron-phonon coupling stabilizes the insulating state. 
The apparent difference
arises from a choice of convention: the authors of Ref.~\cite{Capone04} couple the phonons
to the total density $n$, which implies a $\lambda$-dependent 
shift in chemical potential 
which was interpreted as a physical stabilization. 
We couple the phonons to the difference in density from the half filled
value, which eliminates this (physically irrelevant) shift at half filling.  

We now turn to 
the frequency dependence of the electron self energy.
In weakly correlated materials the 
conduction-band electrons renormalize the phonon propagator $D$ 
from its bare value $D_0$ according to the 
RPA (ladder) result $D^{-1}=D_0^{-1}-\lambda^2\chi_0$
with $\chi_0$ the bare (no phonons) 
electronic density-density correlator.
The Migdal-Eliashberg approximation also implies that the local density-density correlator
$\chi$ is given by
\begin{equation}
\chi(\Omega)=\chi_0(\Omega)/(1-\lambda^2 D_0(\Omega)\chi_0(|\Omega|)).
\label{chirpa}
\end{equation}

The electron self energy is given in frequency space by the convolution  $\Sigma=\lambda^2G_0*D$.
It is convenient to present results as the Matsubara axis mass renormalization factor $r(i \omega_n)=-\Im m\Sigma(i\omega_n)/\omega_n$. 
According to Migdal-Eliashberg theory, the effect of
the electron-phonon coupling is to add to the purely electronic $r(i\omega_n)$ a term 
approximately of the form
\begin{equation}
r_{ME}(i\omega_n)=(2\lambda^2N_0/\omega_n) \arctan(\omega_n/\omega_0^\text{ren})
\label{delZ}
\end{equation}
with $N_0$ the Fermi surface density of states (in our case $\approx 1/2\pi t$). The zero frequency limit gives the electron-phonon contribution to the Fermi surface mass enhancement $m^*_{ME}/m=\frac{2\lambda^2N_0}{\omega_0^\text{ren}}$ 
and $r_{ME}$ drops to about half of its zero frequency value around  $2\omega_0^\text{ren}$.

\begin{figure}[t]
\begin{center}
\includegraphics[angle=-90, width=0.9\columnwidth]{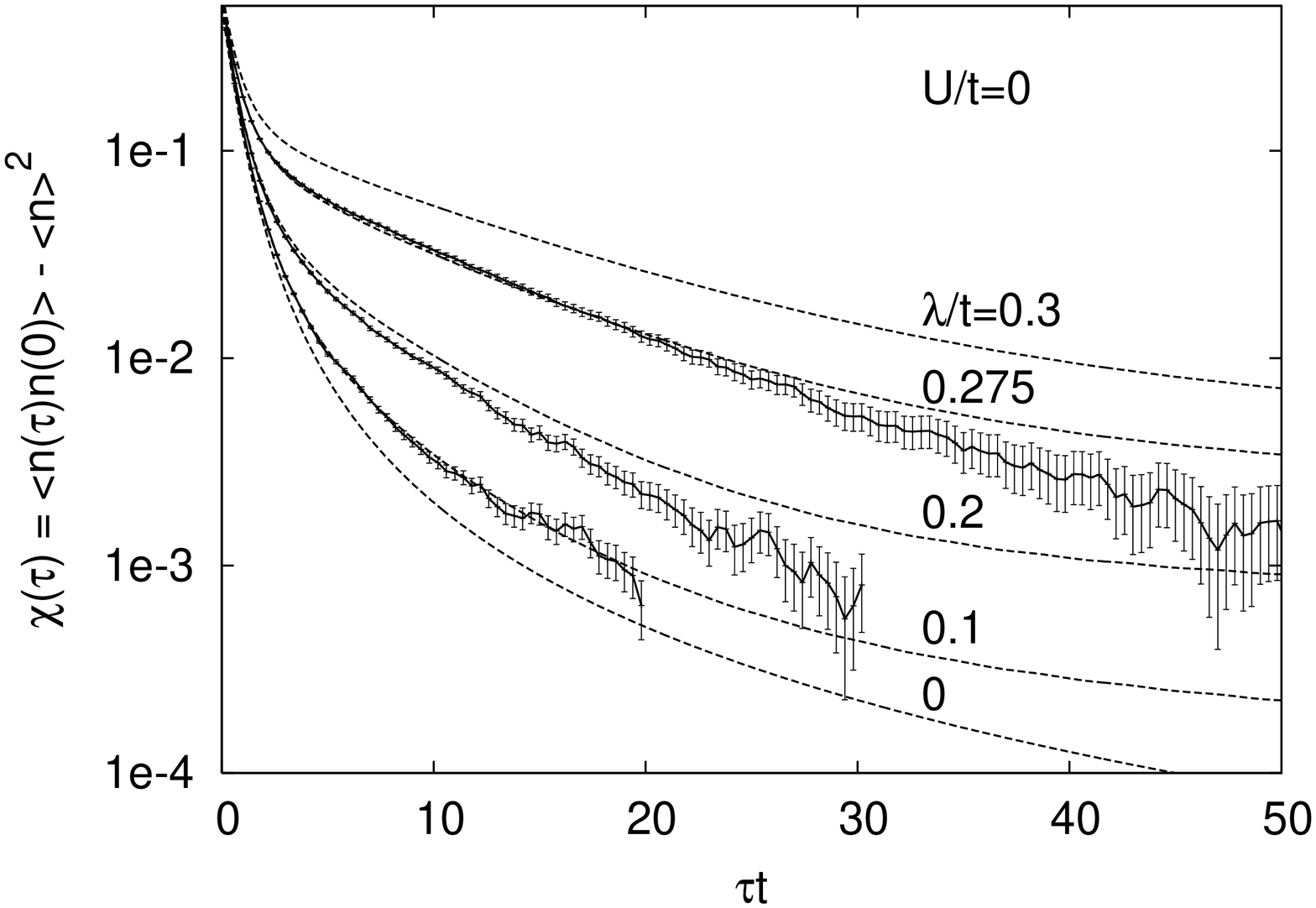}
\includegraphics[angle=-90, width=0.9\columnwidth]{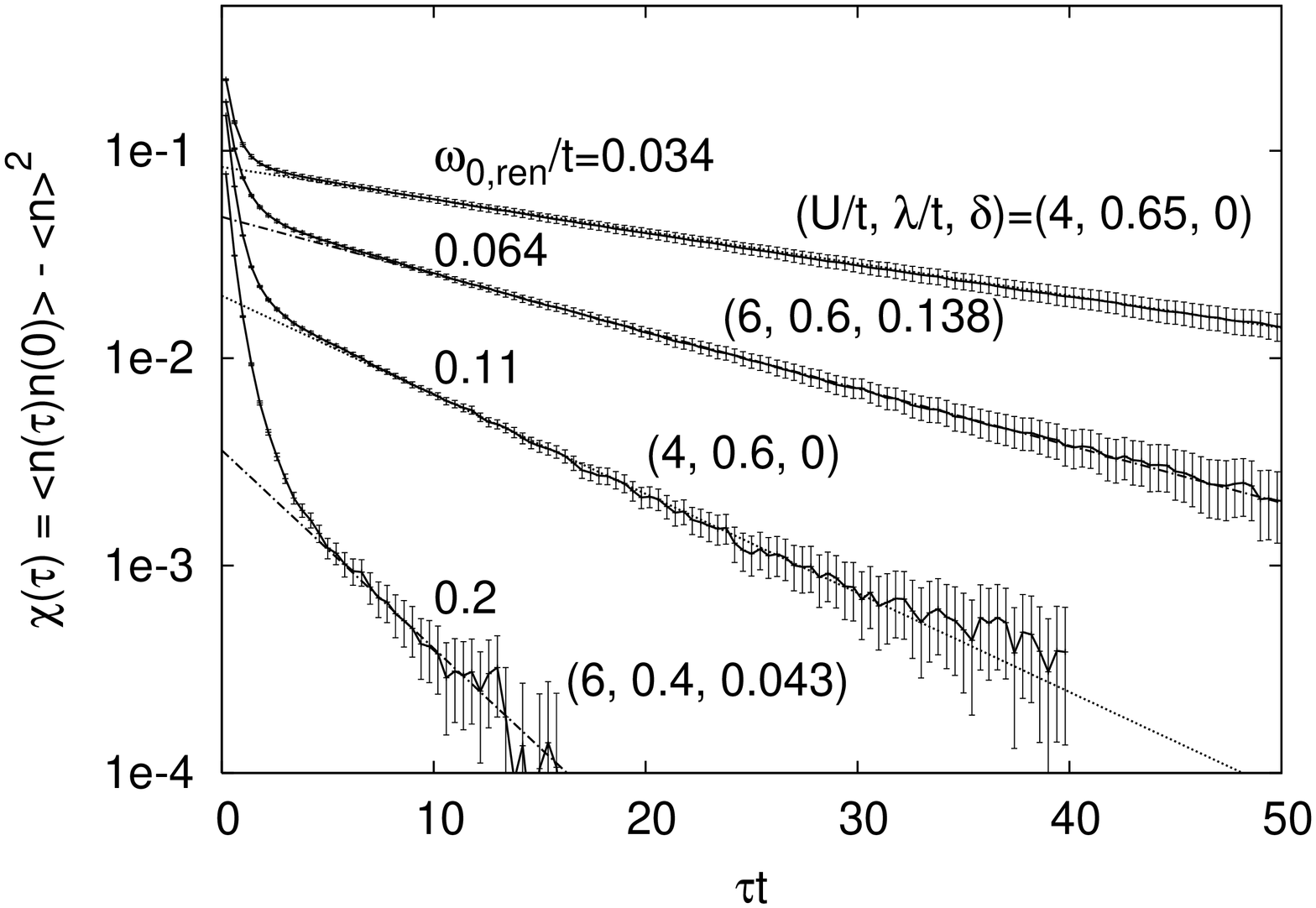}
\caption{Imaginary time dependence of the density-density correlation function plotted on a log-scale to highlight the intermediate time exponential decay. 
If no exponential regime is visible, the phonon frequency is not renormalized; otherwise $\omega_0^\text{ren}$ can be extracted from the slope.
Upper panel: $U=0$ and $n=1$.  Dashed lines: $\chi(\tau)$ from Eq.~(\ref{chirpa}) for indicated phonon couplings. Solid lines with
1-$\sigma$ error bars show the QMC results for $\beta t=400$ 
and $\lambda/t=0.1$, 0.2, 0.3 (from bottom to top).
Lower panel: correlation functions for the interacting model at
half filling, $U/t=4$ and $\lambda/t=0.6$, 0.65  
and in the doped Mott insulator at $U/t=6$, $\lambda/t=0.4$, $0.6$ (dopings per spin $\delta=0.043$, $0.138$).
}
\label{chitauqmcrpa}
\end{center}
\end{figure}
The upper panel of Fig.~\ref{chitauqmcrpa} compares the density-density correlation function computed from Eq.~(\ref{chirpa})  to $\chi(\tau)$ measured in our QMC simulations, for $U=0$ and several values
of $\lambda$. Good agreement is seen; the small shift of $\lambda$ required to match 
the $\lambda=0.3$ QMC data ($\lambda_\text{eff}\approx 0.275$) is a beyond-Migdal 
effect arising from the small but
non-vanishing effect of the electron self energy on $\chi_0$.
$\chi(\tau)$ increases with $\lambda$ because density fluctuations can transform into
phonons. The intermediate-time exponential decay visible in the figure gives the renormalized phonon frequency $\omega^\text{ren}_0$, 
while in the regimes where there is no clear evidence
of exponential decay the phonon frequency is not significantly renormalized.
For $U>0$ the computed $\chi(\tau)$,
shown in the lower panel of Fig.~\ref{chitauqmcrpa}, also  exhibits an exponential decay,
from which we estimate $\omega_0^\text{ren}$. 

\begin{figure}[t]
\begin{center}
\includegraphics[angle=-90, width=0.9\columnwidth]{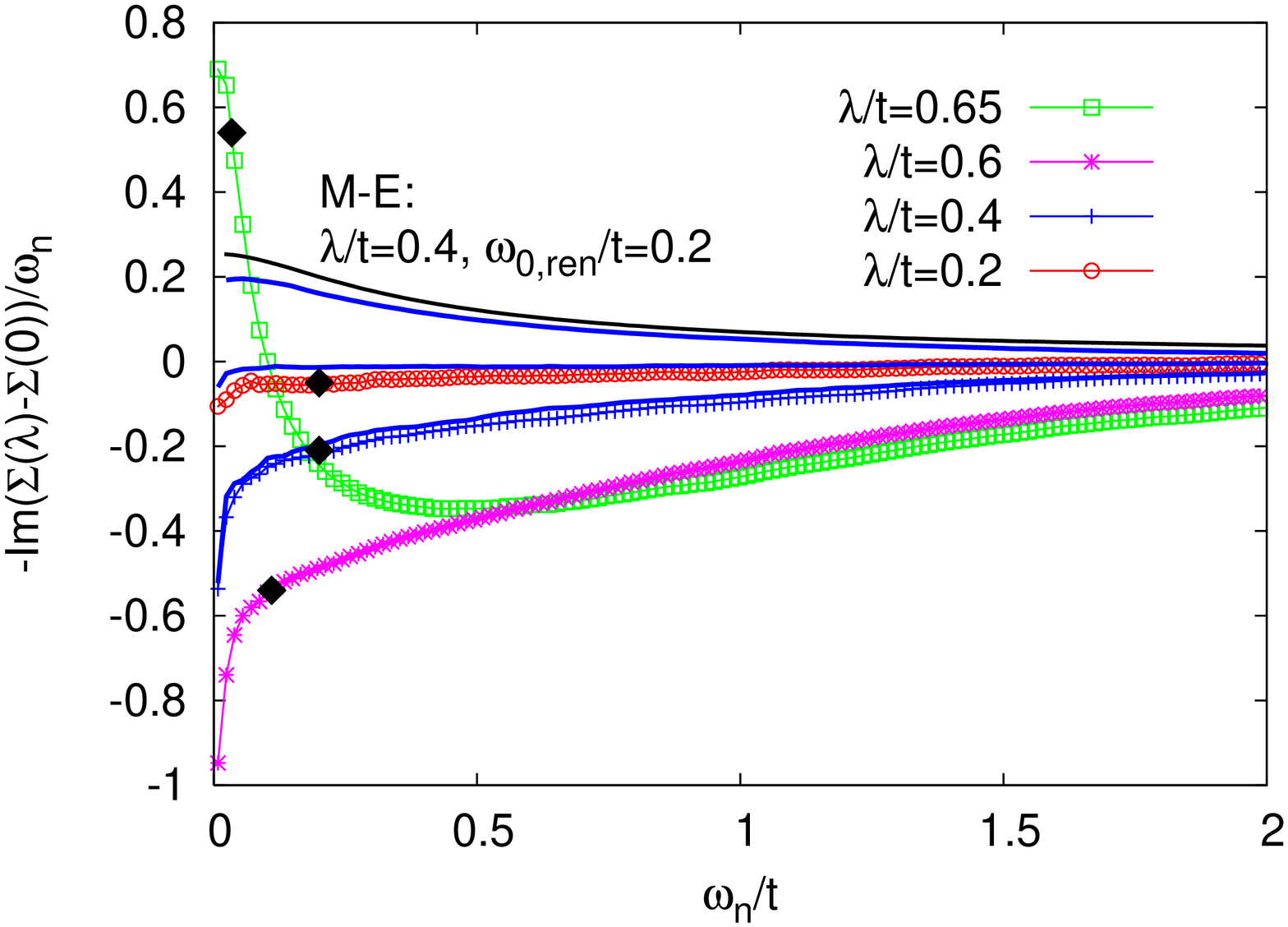}
\includegraphics[angle=-90, width=0.9\columnwidth]{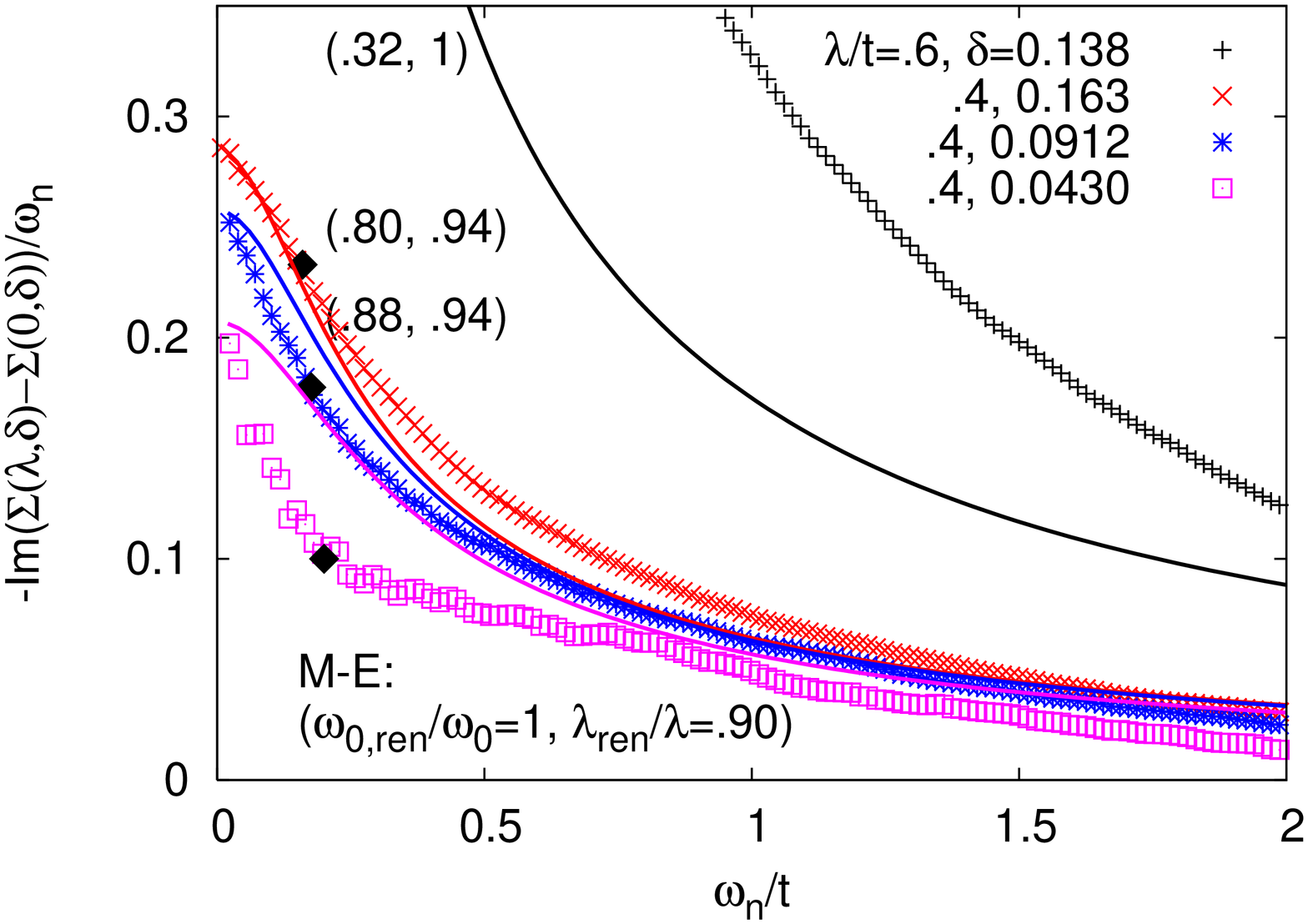}
\caption{
(Color online) Phonon contribution to the electronic self energy 
at $\beta t=400$. Black diamonds:
renormalized phonon frequency ${\omega_0^\text{ren}}$.
Upper panel: $U/t=4$.  Symbols: half filling, $\lambda$ values indicated. Blue
lines: $\lambda/t=0.4$ and doping per spin $\delta=0.019$, 0.067 and 0.182 (from bottom to top).
Black line: Migdal-Eliashberg result.
Lower panel: $U=6t$ (above the critical value for
Mott insulating behavior). Symbols:  $\lambda$ and $\delta$ as shown. Solid lines: Migdal-Eliashberg calculations, Eq.~(\ref{delZ}), with 
renormalized electron phonon couplings $\lambda_\text{ren}$ and phonon frequencies ${\omega_0^\text{ren}}$ as indicated. 
}
\label{self_half_filling}
\end{center}
\end{figure}

Figure~\ref{self_half_filling} presents the phonon-induced change in the mass renormalization function
$\Delta r=-\Im m\left( \Sigma(\lambda,\delta,\omega_n)-\Sigma(0,\delta,\omega_n)\right)/\omega_n$.
The upper panel shows results for $U=4t$, less than the Mott critical value $U_{c2} \approx 5.8t$. 
At half filling (curves traced out by symbols) we find, in agreement with
Ref.~\cite{Koller04}, that the phonon contribution to the mass enhancement has the ``wrong"
sign, except for couplings very close to the bipolaronic instability: the dominant effect of the 
phonons is the reduction of $U$.  A new result is that
(again except for the largest $\lambda$) there is no  obvious feature at the renormalized
phonon frequency. 
The blue lines present the doping dependence at fixed $\lambda=0.4t$. We see
that as the doping is increased, a result compatible with Migdal-Eliashberg theory is eventually recovered. 
The lower panel shows the behavior for $U=6t$, greater than the Mott critical value. Here we find again that for moderate electron phonon couplings and relatively highly doped samples, a renormalized Migdal-Eliashberg theory (with perhaps a slightly reduced effective electron phonon coupling $\lambda_\text{eff}$) 
provides a reasonably consistent description of the data, except that $\Delta r$ drops more rapidly for $\omega_n\gg \omega_0$. However, when the quasiparticle energy $E_F/(1-\partial\Sigma/\partial\omega)$ becomes of order of the phonon frequency (as happens at $\delta=0.04$) or the coupling approaches the bipolaronic instability ($\lambda=0.6t$) the Migdal-Eliashberg description breaks down. For large $\lambda$, $r_{ME}(i\omega_n)$ drops too rapidly, while near the doping driven Mott transition, the shape is too broad. Nevertheless, unlike in the half-filled metal shown in the upper panel of Fig.~\ref{self_half_filling}, in the weakly doped Mott insulator, the upturn in the mass renormalization still coincides with $\omega_0^\text{ren}$. 
Thus, although a Migdal-Eliashberg analysis of self energies in high-$T_c$ materials is problematic, the association of kinks in dispersions with phonon features \cite{Devereaux04} may be robust. 

To summarize: we have introduced a method which enables simulations of impurity models
with electron-phonon coupling at essentially the same computational cost as the corresponding models without phonons. 
We used the method to compute the phase diagram for the doping driven Mott transition in the Holstein-Hubbard model and 
the frequency dependence of the self energy. 
Previous work by one of us and by other groups \cite{Levin,Deppeler02} had argued
on the basis of Fermi liquid theory that the effect of phonons on correlated systems
should be similar to that in uncorrelated systems, but with renormalized couplings.
We found that this expectation is simply wrong at strong interactions in a half-filled model, but applies for large enough fillings and in the doped Mott insulating state as long as the quasi-particle energy is larger than the phonon frequency. 

The approach presented here generalizes 
to other models which 
can be decoupled by Lang-Firsov 
transformations. 
In the Holstein-Hubbard model the interplay
between superconductivity and strong correlations is an important open problem.
Further generalizations are possible, 
for example to Jahn-Teller couplings in multiorbital models -- although
if pair-hopping terms are important one must perform a double-expansion in these
and in the hybridization. Application of the method introduced here to models
of manganites and to $C_{60}$ (if the electronic spectrum is truncated to the
5 levels nearest to the chemical potential) seems entirely feasible. 

The calculations have been performed on the Hreidar beowulf cluster at ETH Z\"urich, using the ALPS-library \cite{ALPS}. We thank M. Troyer for the generous allocation of computer time, K. Ziegler for helpful discussions, and NSF-DMR-040135 for support.

\end{document}